\documentclass[referee,pdflatex,sn-aps]{sn-jnl}

\newcommand{\beq}{\begin{eqnarray}}
\newcommand{\eeq}{\end{eqnarray}}
\usepackage{graphicx}%
\usepackage{multirow}%
\usepackage{amsmath,amssymb,amsfonts}%
\usepackage{amsthm}%
\usepackage{mathrsfs}%
\usepackage[title]{appendix}%
\usepackage{xcolor}%
\usepackage{textcomp}%
\usepackage{manyfoot}%
\usepackage{booktabs}%
\usepackage{algorithm}%
\usepackage{algorithmicx}%
\usepackage{algpseudocode}%
\usepackage{listings}%

\usepackage[12pt]{extsizes}
\usepackage{geometry}
\usepackage[font=small]{caption}

\theoremstyle{thmstyleone}%
\theoremstyle{thmstyletwo}%

\theoremstyle{thmstylethree}%

\begin{document}

\title[Article Title]{Conformally invariant charge fluctuations in a strange metal}

\author*[1,2]{\fnm{Xuefei} \sur{Guo}}\email{xuefeig2@illinois.edu}
\author[1,2]{\fnm{Jin} \sur{Chen}}
\author[1,2]{\fnm{Farzaneh} \sur{Hoveyda-Marashi}}
\author[1,2]{\fnm{Simon L} \sur{Bettler}}
\author[1,2]{\fnm{Dipanjan} \sur{Chaudhuri}}
\author[1,2]{\fnm{Caitlin S} \sur{Kengle}}
\author[3]{\fnm{John A} \sur{Schneeloch}}
\author[3]{\fnm{Ruidan} \sur{Zhong}}
\author[3]{\fnm{Genda} \sur{Gu}}
\author[1,2]{\fnm{Tai-Chang} \sur{Chiang}}
\author[3]{\fnm{Alexei M} \sur{Tsvelik}}
\author[1]{\fnm{Thomas} \sur{Faulkner}}
\author[1,4]{\fnm{Philip W} \sur{Phillips}}
\author*[1,2]{\fnm{Peter} \sur{Abbamonte}}\email{ abbamont@illinois.edu}

\affil*[1]{\orgdiv{Department of Physics}, \orgname{University of Illinois}, \orgaddress{\street{}\city{Urbana}, \postcode{61801}, \state{IL}, \country{USA}}}

\affil*[2]{\orgdiv{Materials Research Laboratory}, \orgname{University of Illinois}, \orgaddress{\street{}\city{Urbana}, \postcode{61801}, \state{IL}, \country{USA}}}

\affil[3]{\orgdiv{Division of Condensed Matter Physics and Materials Science}, \orgname{Brookhaven National Laboratory}, \orgaddress{\street{}\city{Upton}, \postcode{11973}, \state{NY}, \country{USA}}}

\affil[4]{\orgdiv{Institute for Condensed Matter Theory}, \orgname{ University of Illinois}, \orgaddress{\street{}\city{Urbana}, \postcode{61801}, \state{IL}, \country{USA}}}


\maketitle
\textbf{
The strange metal is a peculiar phase of matter in which the electron scattering rate, $\tau^{-1} \sim k_B T/\hbar$, which determines the electrical resistance, is universal across a wide family of materials and determined only by fundamental constants. In 1989, theorists hypothesized that this universality would manifest as scale-invariant behavior in the dynamic charge susceptibility, $\chi''(q,\omega)$. Here, we present momentum-resolved inelastic electron scattering measurements of the strange metal Bi$_2$Sr$_2$CaCu$_2$O$_{8+x}$ showing that the susceptibility has the scale-invariant form $\chi''(q,\omega) = T^{-\nu} f(\omega/T)$, with exponent $\nu = 0.93$. We find the response is consistent with conformal invariance, meaning the dynamics may be thought of as occurring on a circle of radius $1/T$ in imaginary time, characterized by conformal dimension $\Delta = 0.05$. Our study indicates that the strange metal is a universal phenomenon whose properties are not determined by microscopic properties of a particular material. 
}

Simple metals have been in use since the Bronze Age and are among the best understood materials in nature. The current in a metal is carried by electron quasiparticles that scatter as they move, resulting in electrical resistance. The spacing between ions, $a$, sets the minimum possible scattering mean-free path for electrons, $\ell$. As the temperature increases and the mean-free path shortens, the resistivity of a metal eventually saturates when $\ell$ becomes smaller than $a$, a condition known as the Mott-Ioffe-Regel (MIR) limit \cite{Hussey2004}.

The so-called ``strange metals"  exhibit a spectacular violation of the MIR limit \cite{Hussey2004,Ashcroft76}, their resistivity showing a linear dependence on temperature, $T$, including at the lowest temperatures where one expects $\rho \propto T^2$ behavior typical of an ordinary metal \cite{Hussey2004}. 
Other unusual properties include a power-law scaling of the mid-infrared optical conductivity \cite{basov2005electrodynamics,marel2003quantum}, and a single-particle lifetime given by the greater of the energy, $\hbar \omega$, and temperature $T$ \cite{reber2019unified,Vishik2010}. Across a diverse range of strange metals that includes oxides, heavy fermion systems, iron pnictides, organic metals, and more recently, twisted bilayer graphene, a universal feature has emerged: the similar slope of $T$-linear resistivity implies a universal scattering rate determined solely by fundamental constants, $\tau^{-1} \sim k_{\text{B}} T/\hbar$---a phenomenon known as Planckian dissipation. Its consistency across many material families hints at some form of universality that is insensitive to the microscopic details of the material, and was conjectured to represent a universal quantum limit on the current-carrying degrees of freedom a many-body system \cite{zaanen2019planckian}.

Universality in nature is often associated with scale invariance. The best-known example is critical phenomena near a second-order phase transition, whose universal dynamics are governed by general properties such as symmetry and dimensionality than microscopic details of the material \cite{chaikin,goldenfeld}. In 1989, Varma and co-workers 
proposed that the universal properties of strange metals might manifest as scale-invariant behavior in the dynamic charge susceptibility, $\chi''(q,\omega)$ \cite{varma1989phenomenology}. 
Analyzing transport and optical conductivity data in optimally doped cuprates, they conjectured that the susceptibility should be momentum-independent, linear in $\omega$ up to $\omega=k_{\text{B}}T$, with a slope of $1/k_{\text{B}}T$, and then constant-in-frequency for $\omega>k_{\text{B}}T$ (up to a high-energy cutoff) \cite{varma1989phenomenology}.
This conjectured functional form is scale-invariant in that it depends only on the ratio $\omega/T$, meaning the only energy scale is the temperature. We refer to this conjecture as the marginal Fermi liquid (MFL) hypothesis.

An unorthodox feature of the MFL hypothesis is its momentum-independence, which suggests that the critical dynamics in a strange metal are local, meaning excitations do not propagate. 
This peculiar behavior, which resembles that of a zero-dimensional object (e.g., a ``quantum dot"), does not follow from any generally accepted method for computing the charge susceptibility, such as Lindhard theory \cite{PinesNozieres1973}, which predicts a strongly momentum-dependent response \cite{mitrano2018anomalous}. 
The only known microscopic model yielding MFL-like behavior in extended systems is the Sachdev-Ye-Kitaev (SYK) model, which has infinite range interactions specifically designed to exhibit zero-dimensional behavior \cite{SachdevYe1993}
However, relating this model to real materials remains an ongoing challenge \cite{subir1,subir2,subir3}. Despite this, the MFL hypothesis successfully accounts for many key properties of strange metals, including the $T$-linear resistivity, the power-law optical conductivity, and the single-particle lifetime \cite{varma1989phenomenology}. 

The apparent low dimensionality of strange metals points to a possible role for conformal invariance. 
In 1970, Polyakov noted that, in dimensions $d \leq 2$, the correlation functions characterizing critical fluctuations exhibit the full conformal symmetry ---encompassing rotations, dilations, and translations---of which scale invariance is just a specific subset \cite{polyakov1,polyakov2}. In two dimensions the conformal group has an infinite number of generators which imposes stringent constrains on the correlation functions. In particular, it fixes the form of two-point functions independently of the universality class. In (1+1)-dimensional (that is quantum one dimensional) theories, it provides a universal form of two-point correlators at finite temperatures.    

Whether strange metals display scale invariance or full conformal invariance has not been explored experimentally. As noted above, this question hinges on the behavior of the charge susceptibility, $\chi''(q, \omega)$, at large momentum ($q \sim 1/a$) and low energy ($\omega \sim k_{\text{B}} T$) \cite{varma1989phenomenology}. The lack of measurements of the charge susceptibility in this regime is due to the unavailability of momentum-resolved scattering probes that can measure this quantity with milli-electron-volt energy resolution. 

In this paper, we present momentum-resolved electron energy-loss (M-EELS) data for the strange metal Bi$_2$Sr$_2$CaCu$_2$O$_{8+x}$ (Bi-2212) with $T_c = 91$ K. We find that the low-energy density fluctuations, at small $\omega$ and large $q$, exhibit scale invariance, a definitive hallmark of a universal phase. Moreover, these fluctuations show conformal invariance, behaving as though the critical dynamics take place on a circle of radius $1/T$ in imaginary time \cite{Tsvelik1997}. Our study provides evidence that the strange metal phase is a universal, conformally invariant phenomenon, thereby placing strong constraints on the types of microscopic theories suitable for explaining this unusual phase of matter.

Previous M-EELS studies of the high-energy part of the charge response ($\omega>0.1$ eV $\gg k_{\text{B}}T$) in Bi-2212 at small momenta, $q<0.05$ r.l.u. (reciprocal lattice units), revealed a plasmon whose lineshape is consistent with established infrared optics experiments \cite{schulte2002interplay, mitrano2018anomalous,Chen2024}. Plasmons in metals are a well-known consequence of the long-ranged Coulomb interaction, $V(q) = 4\pi e^2/q^2$, which governs the physics at small $q$, in this case $q \lesssim 0.05$ r.l.u. \cite{Chen2024}. Looking at larger momenta, where plasmon effects are less pronounced, Mitrano observed a constant-in-frequency charge response up to 1 eV \cite{mitrano2018anomalous}.  This response was found to be momentum-independent, reminiscent of the MFL hypothesis \cite{varma1989phenomenology}. Husain later showed that this constant-in-frequency behavior was specific to the strange metal part of the phase diagram \cite{husain2019crossover}. However, the most important region of the spectrum, $\omega \sim k_{\text{B}}T$ at large $q$, which would provide insights into the wider claim of scale invariance, remains unexplored. There is therefore a tremendous need to perform M-EELS measurements at nonzero $q$ and at much lower energies $\omega \sim k_{\text{B}} T$.

To experimentally probe the small $\omega$, large $q$ region of the response, we developed a new tuning procedure that improves both the momentum and energy resolution of our M-EELS instrument \cite{Chen2024}. This enhanced setup achieves an energy resolution of 5.5 meV and a momentum resolution of 0.02 \AA$^{-1}$ (see Methods). M-EELS measures the density-density correlation function of the surface of a material, $S(q,\omega)$, also known as the van Hove function \cite{ARCMP}. This function is related to the surface $\chi''(q,\omega)$ through the fluctuation-dissipation theorem, allowing $\chi''(q,\omega)$ to be derived from $S(q,\omega)$ via antisymmetrization \cite{vig2017measurement,ARCMP}.

Using our new tuning method, we recently demonstrated that M-EELS spectra in the optical limit ($q \to 0$) are consistent  with IR optics measurements \cite{Chen2024}. Here, we extend this method to the low-energy, large-$q$ region, where scale-invariant effects might become observable. We focus on the optimally doped strange metal phase of Bi-2212 ($T_c=91$K) grown by floating zone methods \cite{wen2008}. The sample was cleaved {\it in situ} prior to M-EELS measurements, and the raw data were normalized by matrix elements and scaled using a partial $f$-sum rule \cite{vig2017measurement,husain2019crossover} (see Methods). Throughout this paper, momenta are expressed in reciprocal lattice units (r.l.u.) defined in terms of the tetragonal unit cell with $a=3.81$ \AA. All measurements were conducted with $q$ oriented parallel to the (0,1) crystallographic direction unless specified otherwise, where (1,-1) is perpendicular to the supermodulation direction. 

\begin{figure}
    \includegraphics[scale=0.8]{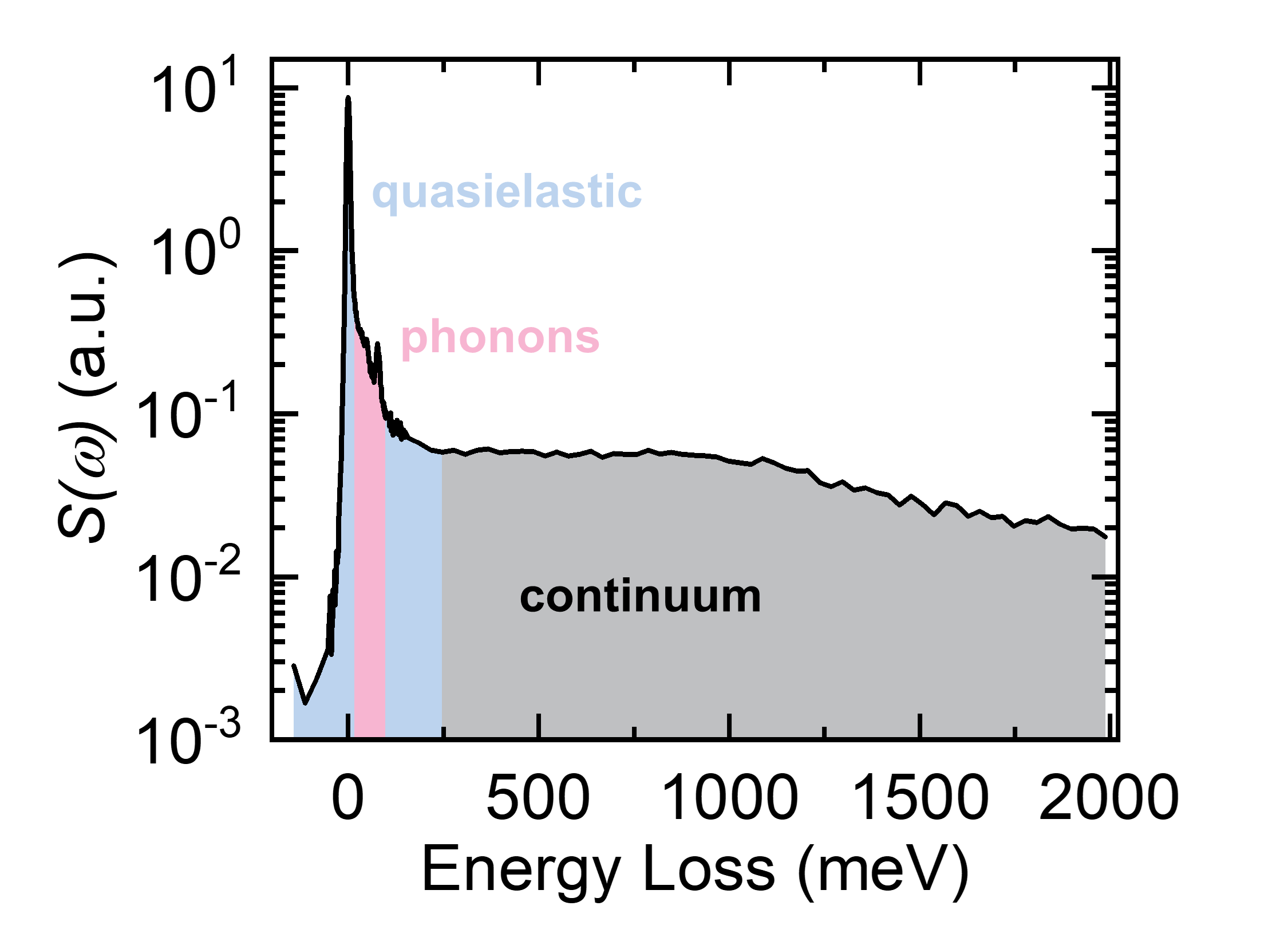}
    \caption{\textbf{The wide energy spectrum at $q\sim0.2$ r.l.u. at 100 K}, showing quasielastic response, phonons and constant-in-frequency continuum.}
    \label{broad}
\end{figure}

A wide M-EELS spectrum taken at a significant in-plane momentum ($q \sim 0.2$ r.l.u.) is shown in Fig. \ref{broad}. The most prominent feature is a frequency-independent, MFL-like continuum, consistent with previous studies \cite{mitrano2018anomalous,husain2019crossover}. This continuum remains constant-in-frequency only within the strange metal part of the phase diagram where the resistivity $\rho \sim T$ \cite{husain2019crossover}. 

In the intermediate energy range (20 meV $< \omega <$ 100 meV), several phonons are visible, consistent with earlier reports \cite{phelps1993,vig2017measurement,persson1990,demuth1990}. At low energy ($\omega <$ 50 meV), a strong quasielastic response, commonly referred to as the ``zero loss line" \cite{Egerton2011}, is observed. 
With the thermal energy scale at room temperature given by $k_{\text{B}} T\sim 25$ meV, the region of interest, $\omega \sim k_{\text{B}} T$, falls within this quasielastic regime.

\begin{figure}
    \centering
    \includegraphics[scale=0.5]{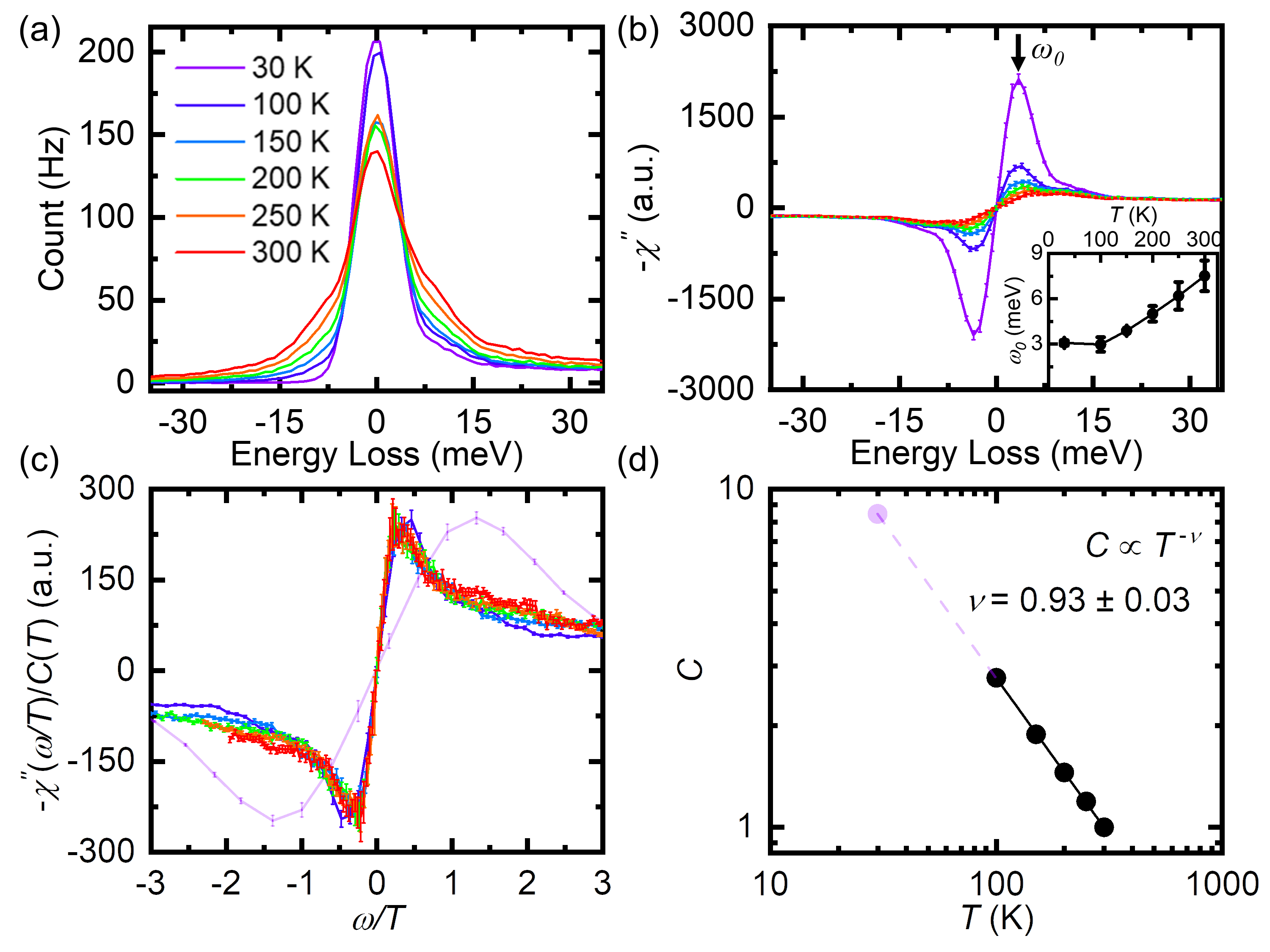}
    \caption{\textbf{Temperature-dependent quasielastic response at $q = 0.36$ r.l.u.} (a) Density fluctuations in the $\omega \sim k_{\text{B}}T $ region, showing that $S(\omega)$ broadens as temperature increases. (b) Antisymmetrized data obtained from (a), where  -$\chi^{\prime\prime}$ crosses zero with a finite slope and exhibits a peak at frequency $\omega_0$. The slope becomes steeper with decreasing temperature, reminiscent of MFL hypothesis. The inset shows the linear-in-$T$ behavior of $\omega_0$, which $\omega_0$ saturates at 3 meV due to the combined effects of superconductivity and the finite energy resolution of the instrument. (c) ``Data collapse” plot, where the susceptibility is plotted against $\omega/T$ and each curve's magnitude is divided by $C(T)$. The data at $T=30$ K is affected by resolution and superconductivity effects. (d) Power-law form of $C(T)$ with $\nu=0.93 \pm 0.03$. This collapse demonstrates that the susceptibility follows the form $\chi^{\prime\prime}(\omega)\propto T^{-\nu}f(\omega/T)$.}
    \label{q_36}
\end{figure}

In Fig. \ref{q_36}(a), we present the temperature dependence of this quasielastic line at a momentum $q$ = 0.36 r.l.u. for 30 K $< T <$ 300 K. Note that the lowest temperature in the spectrum is below $T_c$ with $T_c = 91$ K. The width of the quasielastic line is temperature dependent (Fig. \ref{q_36}(b)), indicating that this feature reflects density fluctuations that are dynamic on an energy scale of $\omega \sim k_{\text{B}}T$. This spectrum represents $S(q,\omega)$, yet the quantity of interest is the susceptibility $\chi''(q,\omega)$. We therefore antisymmetrize $S(q,\omega)$ to obtain $\chi''(q,\omega)$ (Fig. 2(b)). This incurs some errors due to the finite energy resolution, which are discussed in detail in Methods. 

The antisymmetrized data representing $\chi''(q,\omega)$ are shown in Fig. \ref{q_36}(b). The susceptibility passes through $\omega=0$ with a finite slope and shows a peak at a frequency we denote as $\omega_0$ (indicated with an arrow). The slope gets steeper with decreasing temperature, qualitatively similar to the hypothesized $1/k_{\text{B}}T$ slope at $\omega<k_{\text{B}}T$ in the MFL hypothesis \cite{varma1989phenomenology}. The temperature dependence of $\omega_0$ is shown in the inset of Fig. \ref{q_36}(b). We find that the peak energy is approximately proportional to temperature, except at the lowest temperature ($T=30$ K) where $T<T_c$. At this point, $\omega_0$ saturates at 3 meV due to the combined effects of superconductivity and the finite energy resolution of the instrument. This observation suggests that the susceptibility at momentum  $q=0.36$ r.l.u. (excluding the $T=30$ K point, which is influenced by resolution and superconductivity), may be a function only of $\omega/T$, indicative of scale invariance. 

To evaluate this possibility, in Fig. \ref{q_36}(c) we show a ``data collapse”, in which the susceptibility is plotted against $\omega/T$. Additionally, the magnitude of each curve has been scaled by an overall factor, $C(T)$, different for each temperature, which is tuned to minimize the least-squares difference between the curves (see Methods). Except for the $T=30$ K curve, which is affected by resolution and superconductivity effects, the collapse is excellent. The function $C(T)$ follows a power-law form (Fig. \ref{q_36}(d)), $C(T) \propto T^{-\nu}$ with $\nu=0.93 \pm 0.03$. This collapse demonstrates that, in the low energy region within the strange metal phase ($T>T_c$), the susceptibility follows the functional form $\chi''(\omega) \propto T^{-\nu}f(\omega/T)$. This form is characteristic of scale invariance, $T$ being the only energy scale in the data. Note that, while the universal function $f(\omega/T)$ has a constant-in-frequency continuum (Fig. \ref{broad}), reminiscent of the MFL hypothesis, the observed peak at $\omega_0$ was not hypothesized and represents a major departure from MFL as it was originally posed. 

\begin{figure}
    \includegraphics[scale=0.47]{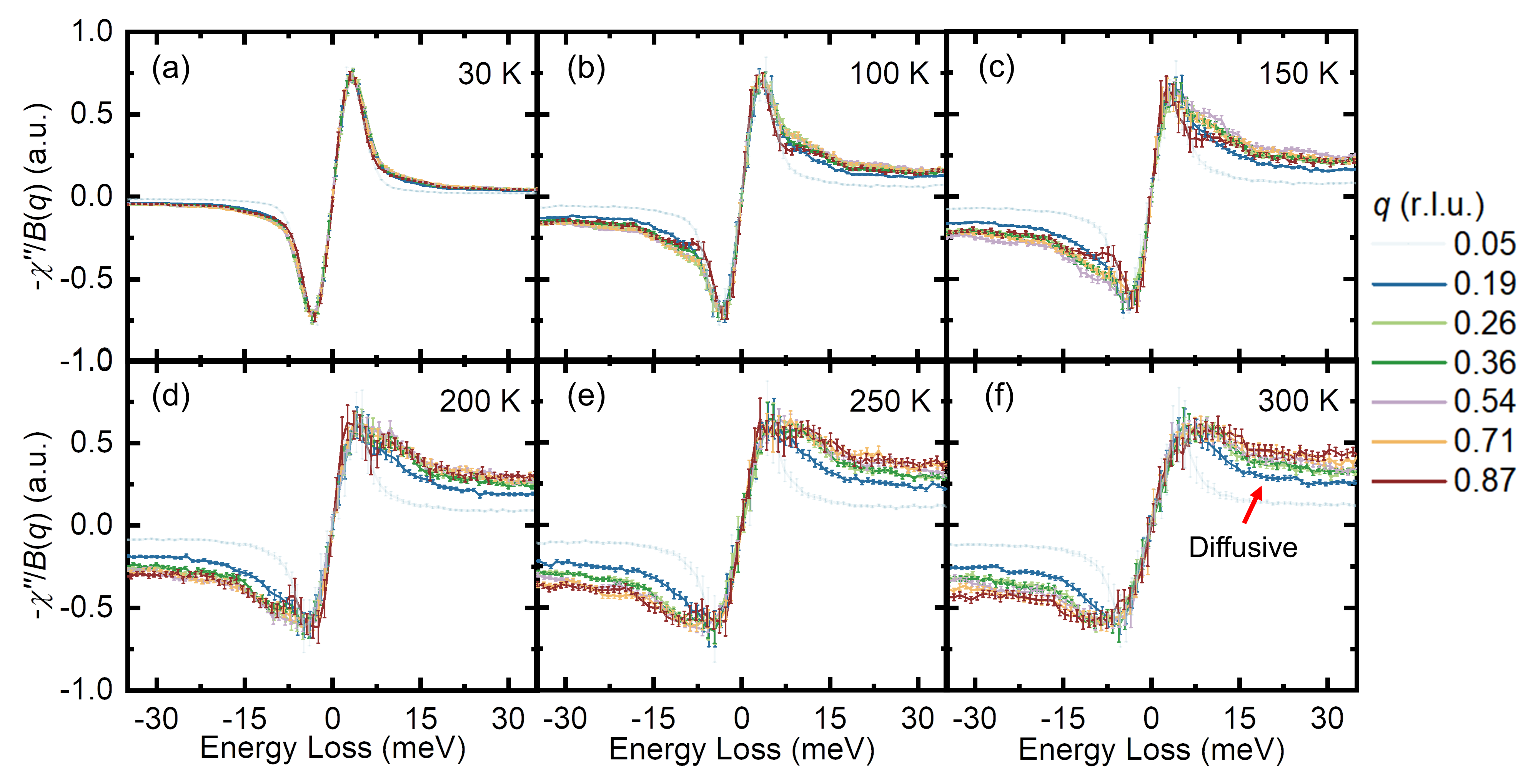}
    \caption{\textbf{Momentum-dependence of susceptibility at different temperatures.} (a)-(f) The susceptibility is divided by $B(q)$ to match the peak height at $\omega_0$ for 30 K $< T < $300 K. The curves exhibit excellent overlap, particularly at $T = 100$ K, indicating that the response is \textit{separable} and can be written as $\chi^{\prime\prime}(q,\omega) \propto B(q)T^{-\nu}f(\omega/T)$. The susceptibility shows increased diffusivity with rising temperature. At $q=0.05$ r.l.u., the data is notably affected by long-ranged Coulomb interactions.}
    \label{momentum}
\end{figure}

We now examine to what extent this behavior depends on the momentum, $q$. In Fig. \ref{momentum}, we show the momentum dependence of $\chi^{\prime\prime}(q,\omega)$ at six different temperatures. Because the data at $q=0.05$ r.l.u. are known to be significantly influenced by long-ranged Coulomb interactions \cite{Chen2024}, we focus on larger momenta, $q\ge 0.19$ r.l.u.. As in Fig. \ref{q_36}(c), we have divided the magnitudes of the curves by an overall factor, $B(q)$, this time to match the peak heights at $\omega_0$ (The unscaled spectra are shown in the Extended Data). 

We first consider the $T=100$ K data (Fig. \ref{momentum}(b)), which is the lowest temperature not obviously influenced by superconductivity or resolution effects (Fig. \ref{q_36}(b), inset). At 100 K, the curves for different $q$ match well, except for the $q=0.05$ r.l.u. spectrum, which is influenced by long-ranged Coulomb interactions \cite{Chen2024}. For larger momenta, $q\ge0.19$ r.l.u., the overlap is nearly perfect. This shows that, for $q\ge0.19$ r.l.u., the susceptibility is \textit{separable}, meaning the response can be written as a product of a function of $q$ and a function of $\omega$, $\chi^{\prime\prime}(q,\omega) \propto B(q)T^{-\nu}f(\omega/T)$ (Note that our experiment shows $B(q)$ to be an increasing function of $q$, but does not strongly constrain its functional form; See Extended Data). This separability was also observed for momenta $q \parallel $ (1,-1) (See Extended Data), suggesting that the effect is isotropic and occurs over $\gtrsim 86\%$ of the Brillouin zone. Separability is an essential feature of the original MFL hypothesis, where it was thought to be a consequence of local scale invariance. Our data at $T=100$ K generally support this hypothesis. 

At higher temperatures, $T>100$ K, the overlap between different momenta is less perfect. A slight momentum dependence of the spectra develops as the temperature approaches room temperature. But, overall, the momentum dependence is weak compared to the propagating excitations observed in EELS measurements on many other materials \cite{Husain2023,Li2023,Kogar2017,Diaconescu2007}, supporting the picture that this three-dimensional system has localized excitations characteristic of a zero-dimensional system. 

\begin{figure}
    \centering
    \includegraphics[scale=0.95]{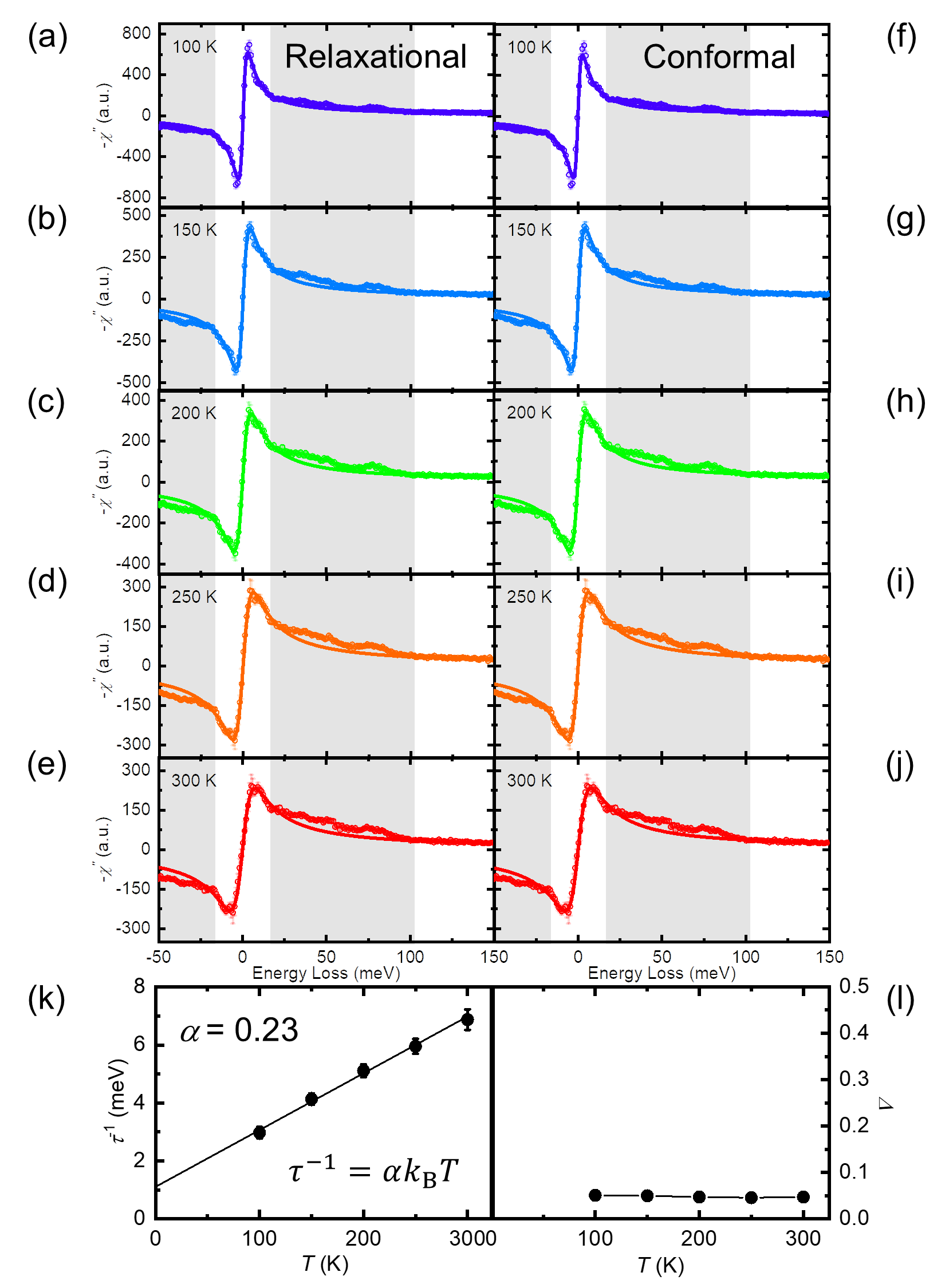}
    \caption{\textbf{Fits of -$\chi^{\prime\prime}$ for different physical models at $q=0.36$ r.l.u.} (a)-(e) Relaxational dynamics of a scalar quantity within the framework of time-dependent Landau theory. Along with observed scale invariance, our data support a quantum critical picture of the density fluctuations in a strange metal. (f)-(j) Conformal invariance of the charge response. The $\omega/T$ scaling and a $2\pi$ relationship between frequency and temperature suggest the existence of conformal invariance in a strange metal. (k) Dissipation in relaxational dynamics shows ``Planckian" behavior with $\alpha = 0.23$. Note that the scattering rate $\tau^{-1}$, representing the dissipation of density fluctuations at finite $q$, differs from the transport scattering rate.  (l) The conformal dimension is temperature-independent with $\Delta = 0.05$. }
    \label{fits}
\end{figure}

Our observations indicate diffusive, relaxational dynamics. The data in Fig. \ref{momentum} are well described by time-dependent Landau theory, specifically ``Model-A" \cite{chaikin}, which characterizes the relaxational dynamics of a scalar quantity \cite{chaikin}. The susceptibility in this model is given by \cite{chaikin}:
\begin{equation}
    \label{relax}
   \chi''(q,\omega,T)=B(q)  \dfrac{\omega}{\omega^2 +\left(\tau^{-1}+Dq^2\right)^2}
\end{equation}
where $\tau^{-1}$ is the relaxation rate and $D$ the diffusion constant. This model accurately fits the susceptibility data at momenta $q\ge0.19$ r.l.u. and temperatures $T\ge100$ K (see Fig. \ref{fits} (a)-(e) and Extended Data). Notably, a good fit was not achieved for $T<T_c$, further supporting the suspicion that spectra at these temperatures are affected by superconductivity and resolution effects. Therefore, our discussion focuses on the strange metal state at $T>T_c$.

Fitting this model provides $\tau^{-1}(T)$ and $D(T)$. The relaxation rate $\tau^{-1}$, shown in Fig. \ref{fits}(k), is linear in temperature, featuring a proportionality constant $\alpha = 0.23$, which is “Planckian” in that $\alpha$ is within an order of magnitude of unity. Notably, the $\tau^{-1}$ measured here, which represents the decay of density fluctuations at finite $q$, has a very different physical meaning from the transport scattering rate. Yet their values are unexpectedly similar. The diffusion constant $D$ is minimal at $T=100$ K but slightly increases at higher temperatures, reflecting the weak momentum dependence in Figs. \ref{momentum}(d)-(f) (see Extended Data). Overall, the dynamics are relaxational, dominated by dissipation at low $T$ but acquiring weak diffusive characteristics as $T$ rises. Relaxational dynamics are typically observed near second-order phase transitions due to critical slowing of fluctuations near a critical point \cite{goldenfeld,chaikin}. This aspect of our data aligns with expectations of critical dynamics, though the presence of a quantum critical point remains uncertain.

The weak momentum dependence shown in Fig. \ref{momentum} is surprising for a three-dimensional material, where one would typically expect propagating excitations \cite{Kogar2017,Husain2023,Diaconescu2007}. This behavior resembles that of a zero-dimensional system.  
However, it remains unclear if this behavior is intrinsic, reflecting true, emergent, local physics, or is merely an experimental artifact. For instance, disorder could induce momentum independence by simply breaking translational symmetry.

To assess this, we checked whether our data are consistent with conformal invariance. As noted earlier, Polyakov showed that systems with critical dynamics in $d<2$ should also exhibit conformal invariance \cite{polyakov1,polyakov2}. Assuming only critical dynamics and conformal symmetry, the charge susceptibility is expected to have the form \cite{polyakov2,faulkner2,Tsvelik1997},
\begin{equation}
\label{CFT}
    \chi^{\prime\prime}(q,\omega,T)=B(q) \; T^{2\Delta-1}\mathrm{Im}\left[\dfrac{\Gamma\left(\Delta-i\omega/\left(2\pi T\right)\right)}{\Gamma\left(1-\Delta-i\omega/\left(2\pi T\right)\right)}\right],
\end{equation}
where $\Delta$ is the conformal dimension and $B(q)$ is a form factor. This expression is, in essence, a statement that the dynamics occur on a circle of radius $1/T$ in imaginary time \cite{Tsvelik1997}. 

We fit our M-EELS data for Bi-2212 in the $\omega \sim k_{\text{B}}T$ region to this expression, focusing on the normal state, where resolution and superconductivity effects are minimal, and on $q \ge 0.19$ r.l.u., where long-range Coulomb interactions are less prominent. The fits are shown in Fig. \ref{fits}(f)-(j) for $q=0.36$ r.l.u., with additional data for other momenta in the Extended Data. The fit quality, assessed using a chi-squared residual, is excellent and comparable to that of relaxational fits. Over the energy range $0 < \omega < 150$ meV, the best fit yields $\Delta = 0.05$, independent of temperature. Notably, the conformal function (Eq. \ref{CFT}) has one fewer fitting parameter than the relaxational function (Eq. \ref{relax}), as the conformal dimension appears in both the temperature prefactor and the argument of the $\Gamma$-functions. 
Additionally, the $2 \pi$ factor in the $\Gamma$-function arguments was found to be essential; omitting it results in a poorer fit (see Extended Data). We conclude that 
the local character of density fluctuations we observe in Bi-2212 is intrinsic, and not due to a disorder effect. Further, the dynamics are conformally invariant, the key experimental signatures being $\omega/T$ scaling and a $2 \pi$ relationship between frequency and temperature.

Our measurements suggest a subtle connection between relaxational dynamics and holographic descriptions of strange metals. The charge dynamics we observe in Bi-2212 are scale-invariant and fit equally well by a Model-A description of relaxational dynamics, Eq. \ref{relax}, and by Eq. \ref{CFT}, which assumes only conformal invariance. In Methods, we show that the latter expression reduces to the former in a suitable limit. Two-point correlation functions identical to Eq. \ref{CFT} appear in holographic models of finite-density matter in (d+1)-dimensional anti-de Sitter (AdS$_{d+1}$) spacetime, where the IR limit is governed by a 
(0+1)-dimensional (AdS$_2\times R^d$) conformal theory \cite{faulkner2,faulkner1}. Conformal invariance reflects the metric space of the underlying theory and offers a pathway to constructing the higher-dimensional framework. Our findings suggest a deep link between holographic constructions, conformal invariance, and the physical behavior of relaxational dynamics.

While our findings show notable similarities with the Marginal Fermi Liquid (MFL) hypothesis from 1989, there are key differences. Most importantly, we observe a peak at $\omega \sim k_{\text{B}}T$, which is absent in the MFL hypothesis. MFL corresponds to Eq. \ref{CFT} with a conformal dimension $\Delta = 1/2$ at which density fluctuation peaks are suppressed \cite{faulkner2}. In the current experiment, we find a conformal dimension $\Delta = 0.05$ associated with a pronounced peak in the response at $\omega \sim k_{\text{B}}T$. 
Because most strange metals are quasi-2D materials, these results suggest  suggesting a decoupling similar to the AdS$_{2}\times R^d$ reduction. Such a decoupling may arise in a broader class of microscopic theories. Whatever the final theory entails, our experimental findings highlight conformal invariance as a central feature of strange metal physics, regardless of its spatial dimensionality \cite{Gori3dCI,Rattazzi}.

{\bf Acknowledgements.} 
We acknowledge G. Kotliar for helpful discussions. 
This work was primarily supported by the Center for Quantum Sensing and Quantum Materials, an Energy Frontier Research Center funded by the U.S. Department of Energy (DOE), Office of Science, Basic Energy Sciences (BES), under award DE-SC0021238. Growth of Bi-2212 single crystals was supported by DOE Grant No. DE-SC0012704. P.A. gratefully acknowledges additional support from the EPiQS program of the Gordon and Betty Moore Foundation, grant GBMF9452. 
A.M.T. acknowledges support from DOE BES Contract No. DE-SC0012704.

\newpage

\backmatter

\section*{Methods}
\subsection*{M-EELS measurements}
The monochromatic beam is tuned in the direct beam configuration. Voltages on the lenses and monochromators are systematically optimized according to the energy loss-angle map, to achieve a balance between energy resolution, momentum resolution and beam current \cite{Chen2024}. The energy resolution is estimated by integrating over the angle from the phase space map, where the energy profile is accurately represented by a Gaussian fit. The 5.5 meV energy resolution is further confirmed by the elastic line at the specular reflection of Bi-2212 at 30 K, showing a consistency with the direct beam estimate within 0.1 meV.

M-EELS spectra are averaged and then divided by the matrix element to obtain $S(\omega)$. The quantity of interest $\chi^{\prime\prime}(\omega)$ is related to $S(\omega)$ through the fluctuation-dissipation theorem \cite{vig2017measurement}. The response function, labeled as $\chi^{\prime\prime}_B(\omega)$ and $\chi^{\prime\prime}_A(\omega)$, is derived from $S(\omega)$ either by dividing out the Bose factor or by antisymmetrization, respectively. To determine the energy shift $\omega_s$ of the zero of the energy loss, we minimize the least-squares difference between $\chi^{\prime\prime}_B(\omega-\omega_s)$ and $\chi^{\prime\prime}_A(\omega-\omega_s)$. The final $\chi^{\prime\prime}(\omega)$ is taken as the average of $\chi^{\prime\prime}_B(\omega)$ and $\chi^{\prime\prime}_A(\omega)$ after applying the energy shift $\omega_s$. The difference between these two values provides the error estimation, and the error from the Poisson statistics of the signal is propagated through $\chi^{\prime\prime}_B(\omega)$ and $\chi^{\prime\prime}_A(\omega)$. The energy gain portion of the spectra is then plotted from the antisymmetrized energy loss data. The uncertainty arising from this procedure to obtain $\chi^{\prime\prime}(\omega)$ is attributed to the finite energy resolution of our instrument.

M-EELS spectra follow the $f$-sum rule \cite{Husain2023},
\begin{equation}
\label{sum-rule}
    \int_0^{\infty}\omega\chi^{\prime\prime}(q,\omega)\mathrm{d}\omega=-\pi\frac{\hbar^2q^2}{2m}\int_{-\infty}^{0}\rho(z)e^{2qz}\mathrm{d}z,
\end{equation}
where $m$ is the electron mass, $q$ is the momentum and $\rho(z)$ is the material density along $z$, perpendicular to the cleavage surface. Assuming 
\begin{equation}
\label{charge_dist}
    \rho(z)=\rho_0\left(1-e^{z/d}\right),
\end{equation}
where $d$ is the depth of the surface layer, the sum rule simplifies to $\dfrac{kq}{2qd+1}$ with $k=-\dfrac{\pi\rho_0\hbar^2}{4m}$. In this work, we apply partial sum rule by setting the upper limit of integration to 150 meV rather than extending to infinity, though this approach remains unproven in strange metals. Furthermore, the functional form of the sum rule depends strongly on the assumed value of $d$; for simplicity, we set $d=0$ throughout the analysis unless otherwise specified. All $\chi^{\prime\prime}(\omega)$ values in this work are scaled using this partial $f$-sum rule.

\subsection*{$\omega/T$ ``data collapse”}
The energy range for performing the ``data collapse” is set within $|\omega/k_{\text{B}}T|<1$, excluding the phonon parts of the spectra. To determine the scaling factor $C(T)$, we minimize the least-squares difference of $\chi^{\prime\prime}(\omega/T)/C(T)-\dfrac{1}{N}\sum\limits_{T}\chi^{\prime\prime}(\omega/T)/C(T)$ for all temperatures and within the specified energy range, incorporating the uncertainty in $\chi^{\prime\prime}$. Here, $N$ is the total number of the curves for different temperatures included in the analysis.

\subsection*{Fitting of the relaxational and conformal model}
Both the relaxational and conformal model fits exclude the phonon regions of the spectra and do not use the uncertainty in $\chi^{\prime\prime}$ as weights. To prevent bias towards low-temperature spectra, $\chi^{\prime\prime}$ is normalized to 1 at each temperature. At each $q$, a global fit is performed across all temperatures above $T_c$, with $A_R$ in Eq. \ref{relax} and $A_C$ in Eq. \ref{CFT} treated as a shared parameter within each model. In the relaxational model, we obtain $\gamma(q)=\tau^{-1}+Dq^2$ for each $q$ and fit $\gamma(q)$ to this parabolic form to determine the relaxation rate $\tau^{-1}$ and the diffusion constant $D$. For the conformal model, note that $\Delta=0.05$ describes the data well up to 150 meV but does not work at higher energies. To achieve a constant-in-frequency form at higher energies, $\Delta=0.5$ is required.

\subsection*{Mapping between the relaxational and conformal model}
The question arises as to why relaxational and conformal responses coincide.  This can be seen from the simple argument based on the identity: $\Gamma(1-z)\Gamma(z)=\pi/\sin\pi z$.  With this identity in tow, we rewrite $1/\Gamma(1-\Delta-ix)=\Gamma(\Delta+ix)\sin(\pi(\Delta+ix)/\pi$.  Consequently, 
\beq
{\rm Im} \frac{\Gamma(\Delta-ix)}{\Gamma(1-\Delta-ix)}=\frac{|\Gamma(\Delta-ix)|^2}{\pi}\cos\Delta\sinh x.
\eeq
Noting that in this problem $x=\omega/T$ and $\sinh x\propto x$ for small values of $x$, we find that $\chi^{''}=(\omega/T) |\Gamma(\Delta-ix)|^2\cos\Delta$. Making contact with relaxational dynamics is now immediate because $\Gamma(\Delta-ix)$ is dominated by its pole at $\Delta-ix=0$.  This allows us to replace $|\Gamma(\Delta-ix)|^2$ by $1/(\Delta^2+x^2)$.  This results in precisely the form of Eq. \ref{CFT} implying the consilience of relaxational dynamics and conformal invariance at least in the limited parameter regime where $\Gamma(y)\approx 1/y$.  Notably, the spin susceptibility in the heavy fermion UCu$_{5-x}$Pd$_x$\cite{Tsvelik1997} also obeys conformal invariance as dictated by Eq. \ref{CFT}.  The underlying rationale here is the inherent conformal structure\cite{Tsvelik1997,polchinski} of the (0+1)-dimensional impurity models invoked to explain heavy fermions.

\section*{Extended Data}
\setcounter{figure}{0}
\renewcommand{\figurename}{Extended Data Fig.}
\renewcommand{\thefigure}{\arabic{figure}}

\begin{figure}[H]
    \centering
    \includegraphics[scale=0.47]{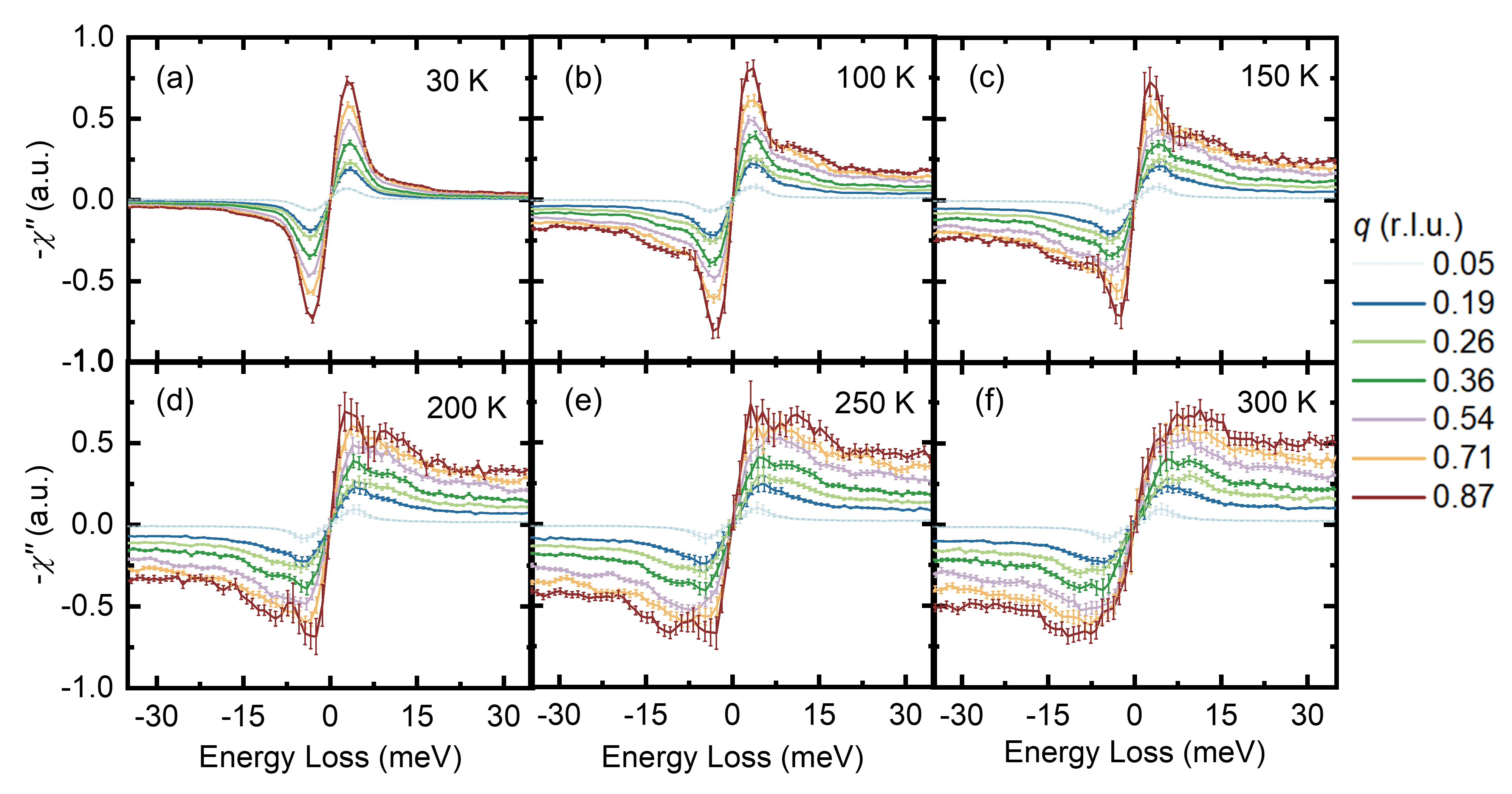}
    \caption{\textbf{Unscaled momentum-dependence of susceptibility at different temperatures.} (a)-(f) The peak shape and position of the susceptibility are highly similar for different momenta at the same temperature, especially at $T = 100$ K, suggesting that the charge response $\chi^{\prime\prime}$ is \textit{separable}.}
    \label{momentum_unscaled}
\end{figure}

\begin{figure}[H]
    \centering
    \includegraphics[scale=0.47]{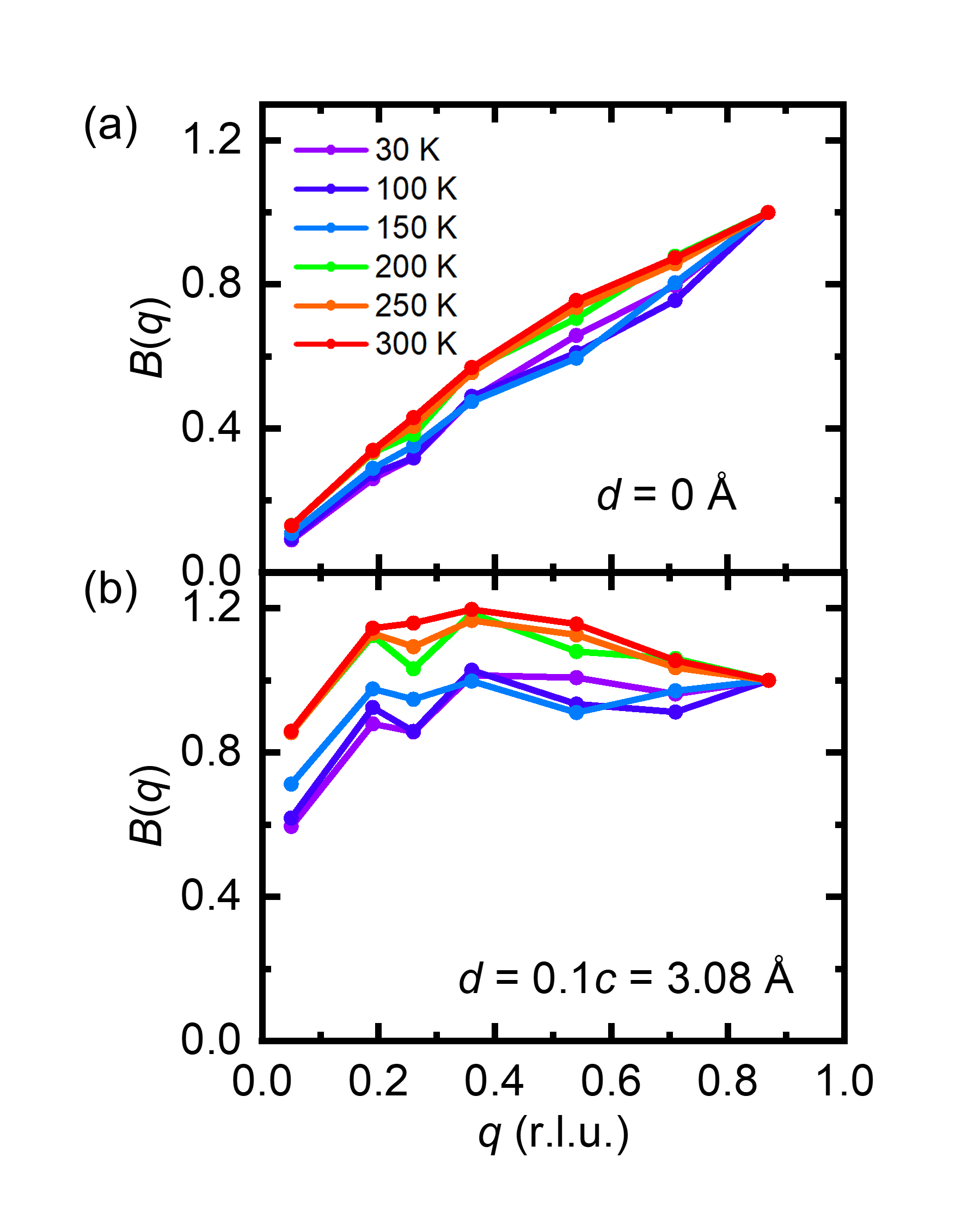}
    \caption{\textbf{Scaling factor $B(q)$ in Fig. \ref{momentum} demonstrating ``local scale invariance”.} (a,b) $B(q)$ strongly depends on the depth of the surface layer $d$ in Eq. \ref{charge_dist}, but it overall is a monotonically increasing function of $q$.}
    \label{B_q}
\end{figure}

\begin{figure}[H]
    \centering
    \includegraphics[scale=0.47]{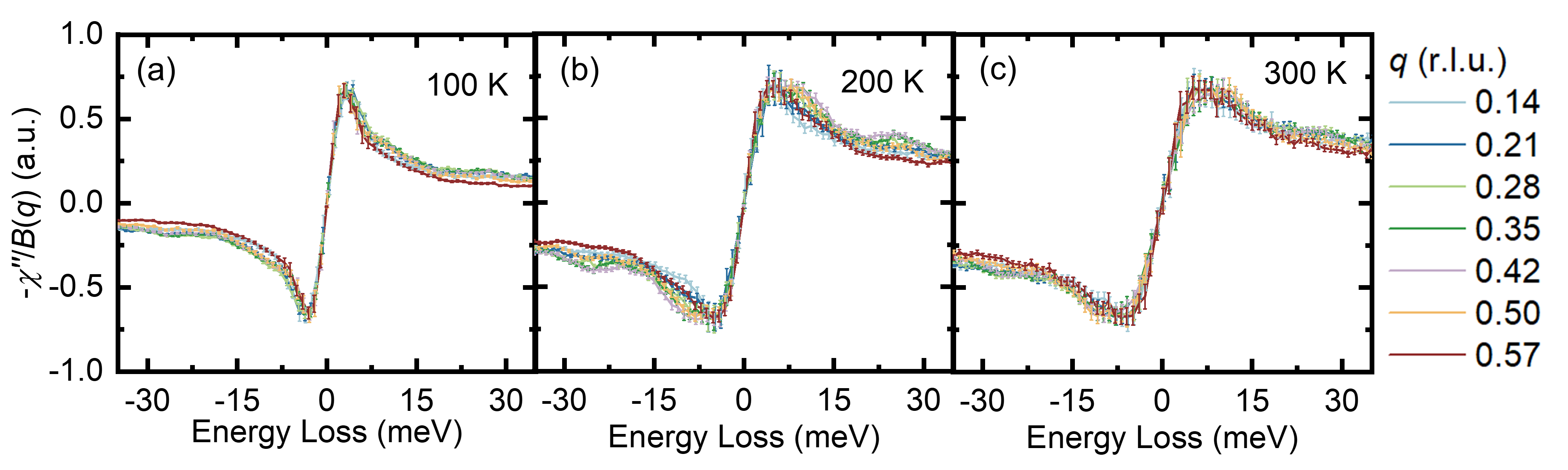}
    \caption{\textbf{Momentum-dependence of susceptibility at different temperatures for $q \parallel $ (1,-1).} (a)-(c) The susceptibility is divided by $B(q)$ to match the peak height at $\omega_0$ for 100 K $< T < $300 K. The curves exhibit excellent overlap, suggesting that the \textit{separability} of the charge response is isotropic.}
    \label{momentum_11}
\end{figure}

\begin{figure}[H]
    \centering
    \includegraphics[scale=0.49]{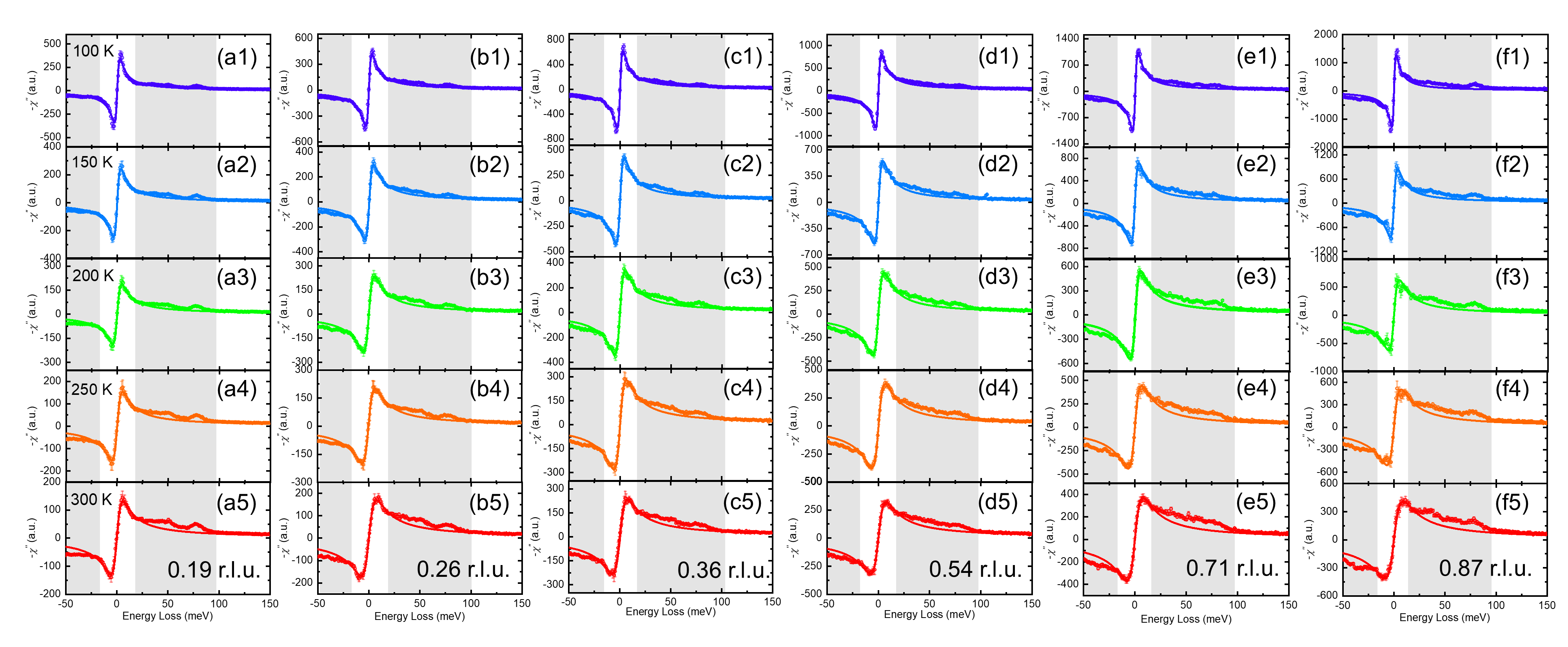}
    \caption{\textbf{Fits of the relaxational model in Eq. \ref{relax}.} The relaxational model fits the data above $T_c$ reasonably well at (a1)-(a5) $q = $ 0.19 r.l.u., (b1)-(b5) 0.26 r.l.u., (c1)-(c5) 0.36 r.l.u., (d1)-(d5) 0.54 r.l.u., (e1)-(e5) 0.71 r.l.u., and (f1)-(f5) 0.87 r.l.u.. The gray-shaded regions are the phonon parts of the spectra and are not included in the data fitting.}
    \label{relax_fits}
\end{figure}

\begin{figure}[H]
    \centering
    \includegraphics[scale=0.6]{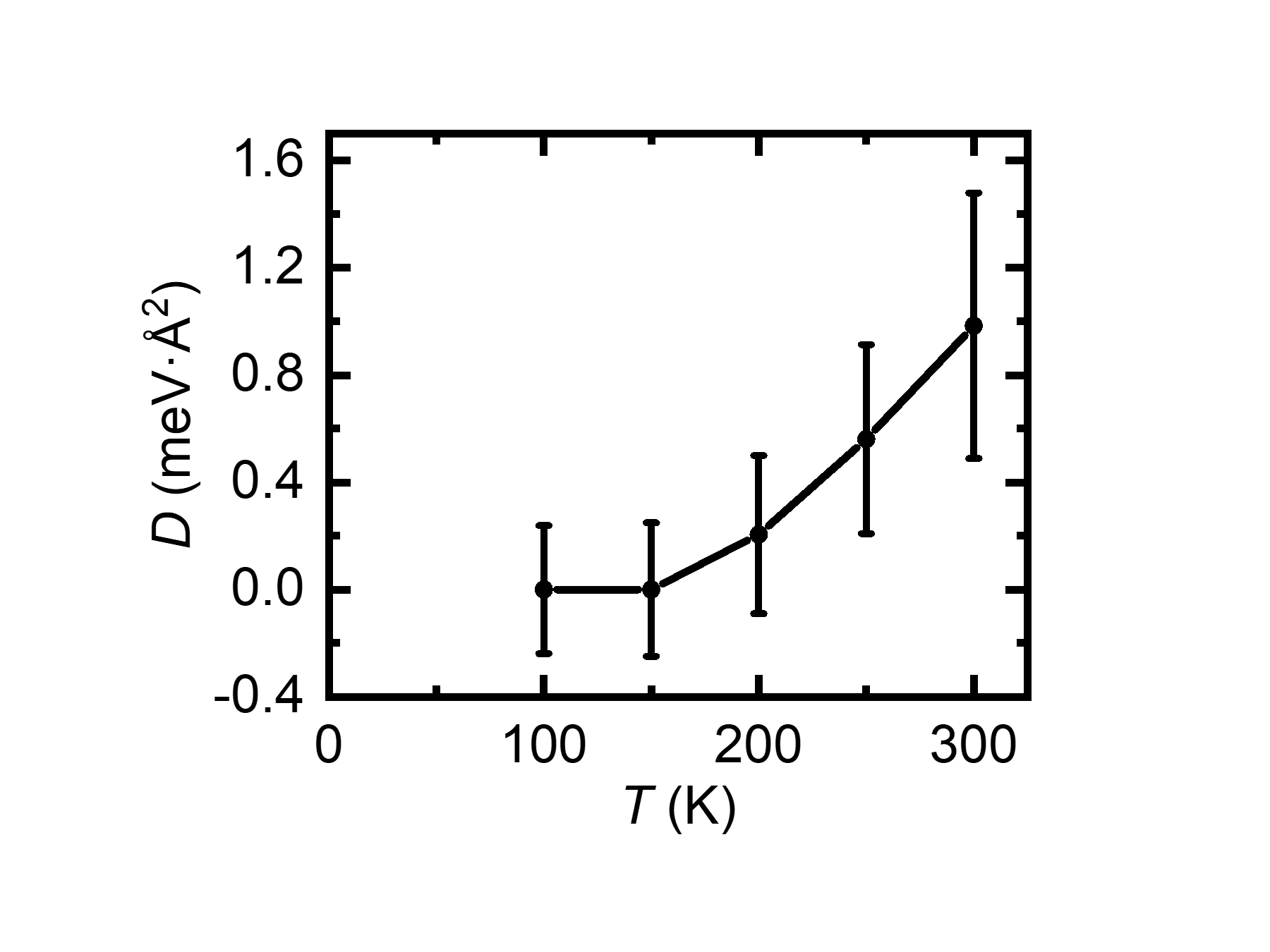}
    \caption{\textbf{Temperature dependence of the diffusion constant.} The susceptibility begins to show diffusive behavior at higher temperatures.}
    \label{relax_D}
\end{figure}

\begin{figure}[H]
    \centering
    \includegraphics[scale=0.95]{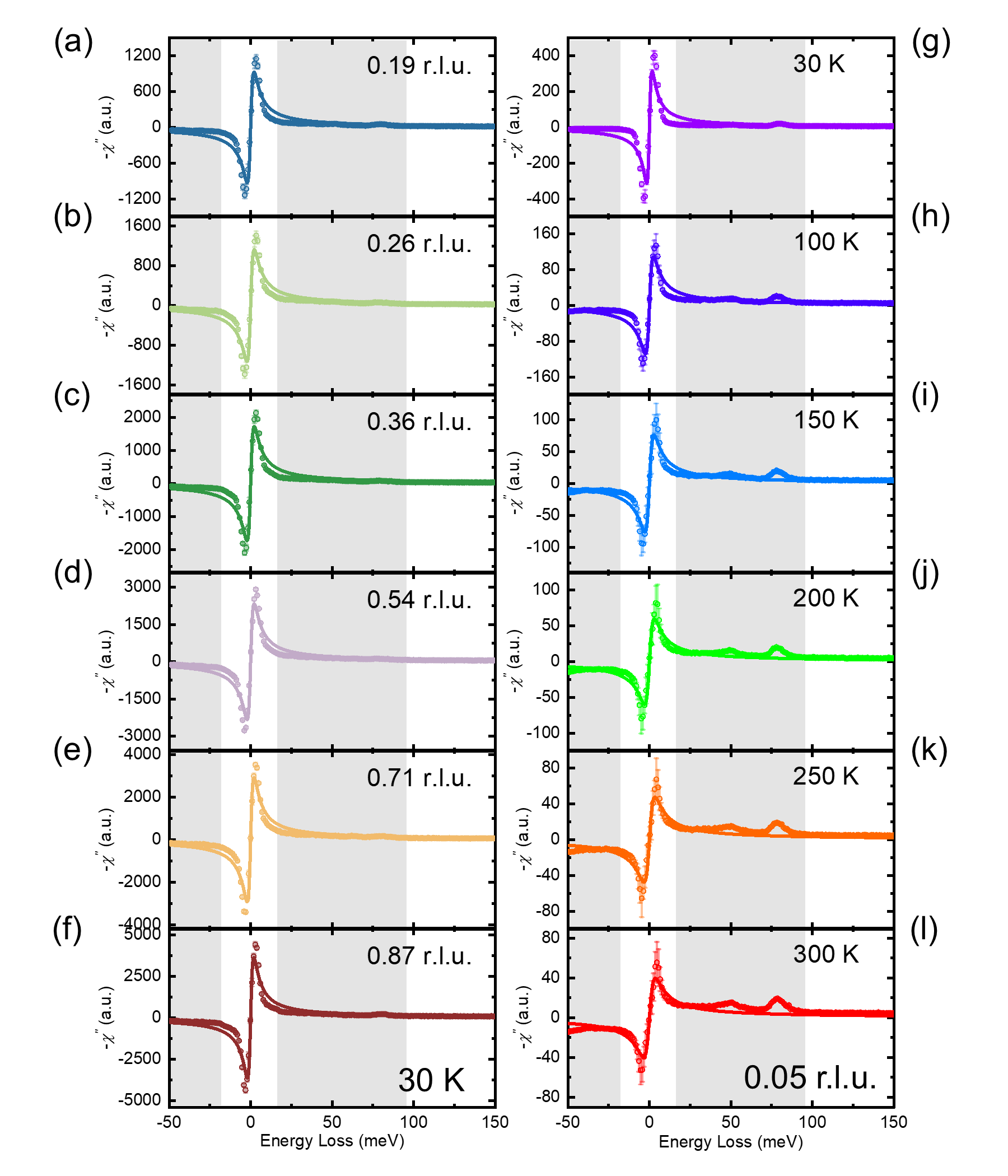}
    \caption{\textbf{Relaxational fits at $T = $ 30 K ($T<T_c$) and $q=0.05$ r.l.u..} (a)-(f) The relaxational model fits the data poorly at $T=30$ K, where the spectra are affected by superconductivity or resolution effects. (g)-(l) The relaxational model fails to capture the dynamics at $q=0.05$ r.l.u., where the spectra are influenced by long-ranged Coulomb interactions.}
    \label{relax_poor_fits}
\end{figure}

\begin{figure}[H]
    \centering
    \includegraphics[scale=0.49]{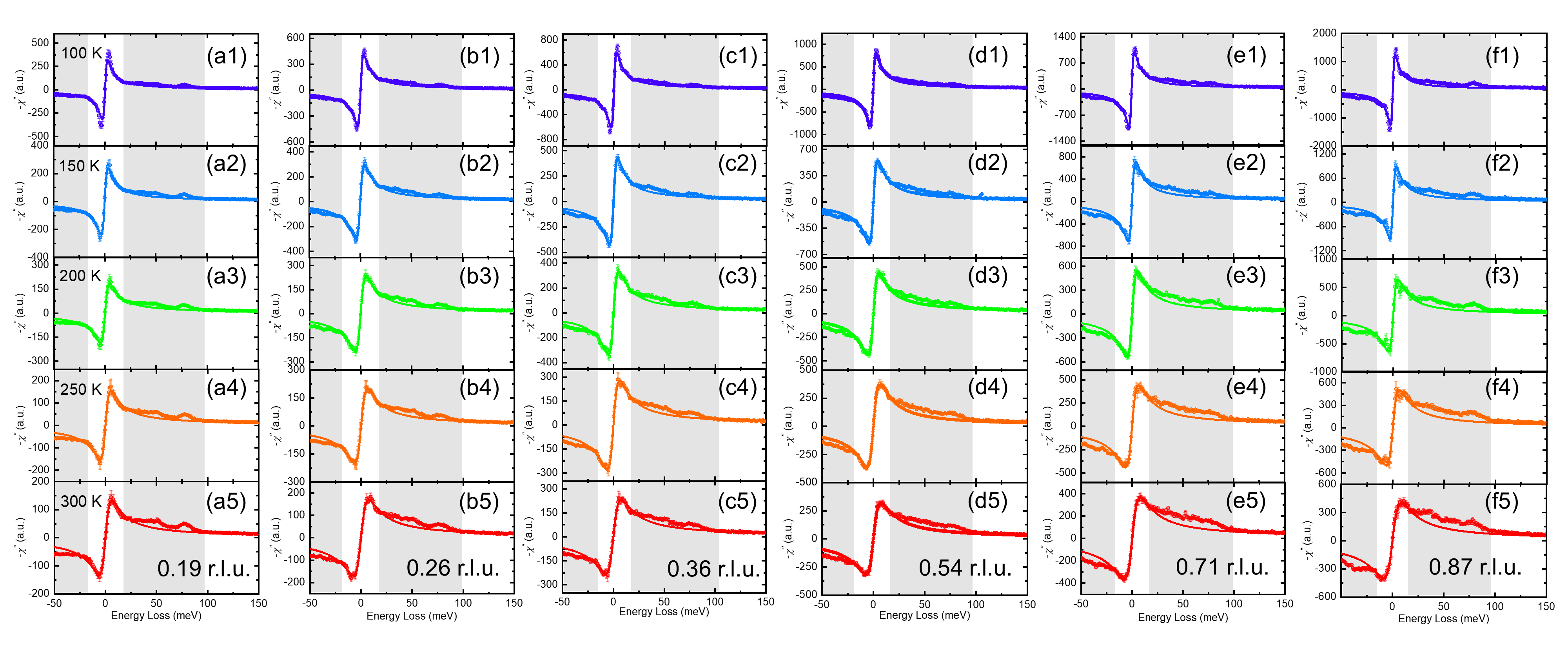}
    \caption{\textbf{Fits of the conformal model in Eq. \ref{CFT}.} The conformal model fits the data above $T_c$ reasonably well at (a1)-(a5) $q = $ 0.19 r.l.u., (b1)-(b5) 0.26 r.l.u., (c1)-(c5) 0.36 r.l.u., (d1)-(d5) 0.54 r.l.u., (e1)-(e5) 0.71 r.l.u., and (f1)-(f5) 0.87 r.l.u.. The gray-shaded regions are the phonon parts of the spectra and are not included in the data fitting.}
    \label{CFT_fits}
\end{figure}

\begin{figure}[H]
    \centering
    \includegraphics[scale=0.95]{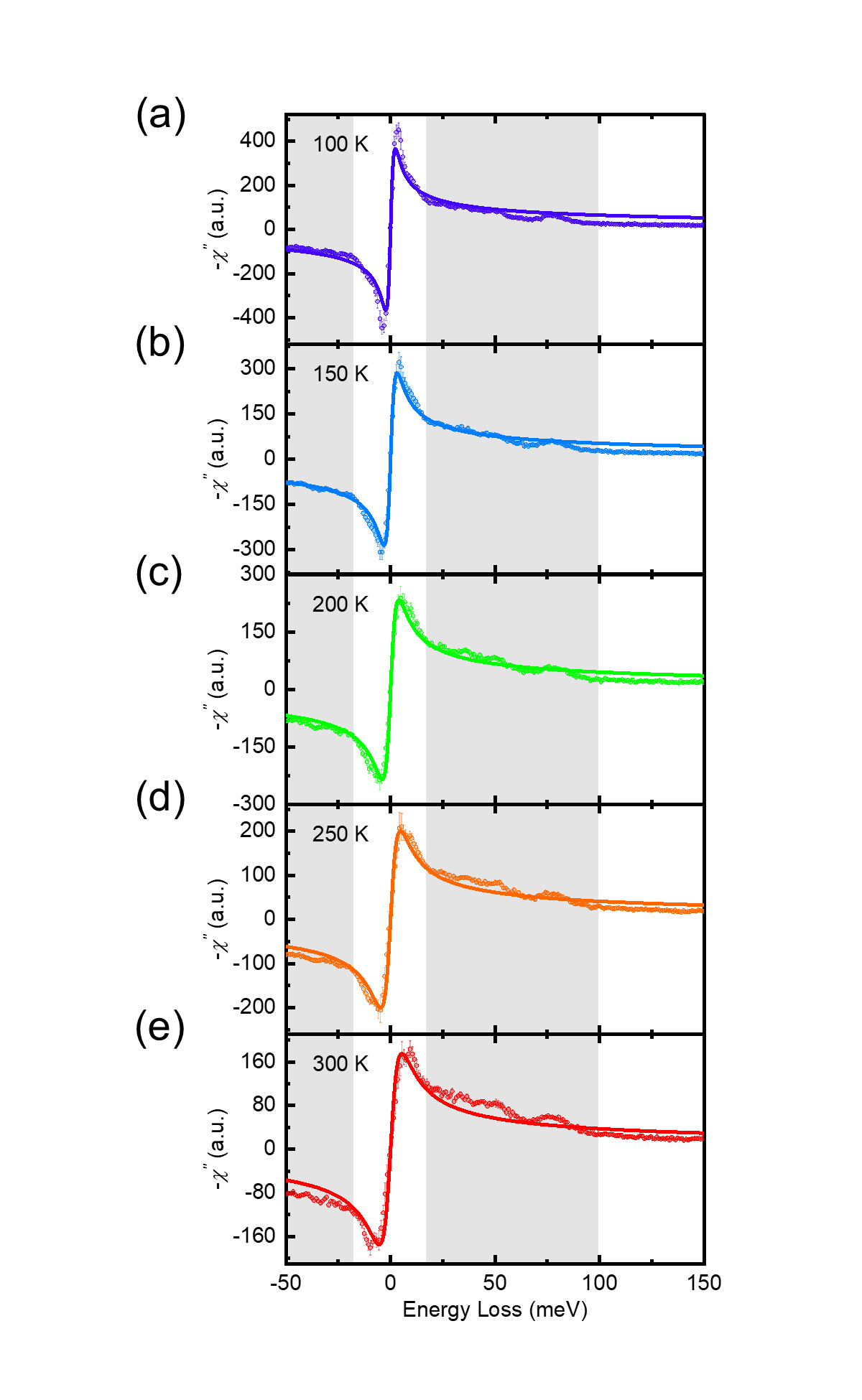}
    \caption{\textbf{$2\pi$ relationship between frequency and temperature.} (a)-(e) Fits of the conformal model without the $2\pi$ factor in the arguments of the $\Gamma$-function in Eq. \ref{CFT} at $q=0.26$ r.l.u.. Achieving a good fit requires including the $2\pi$ factor, as shown in Extended Data Fig. \ref{CFT_fits}}
    \label{CFT_2pi}
\end{figure}

\begin{figure}[H]
    \centering
    \includegraphics[scale=0.5]{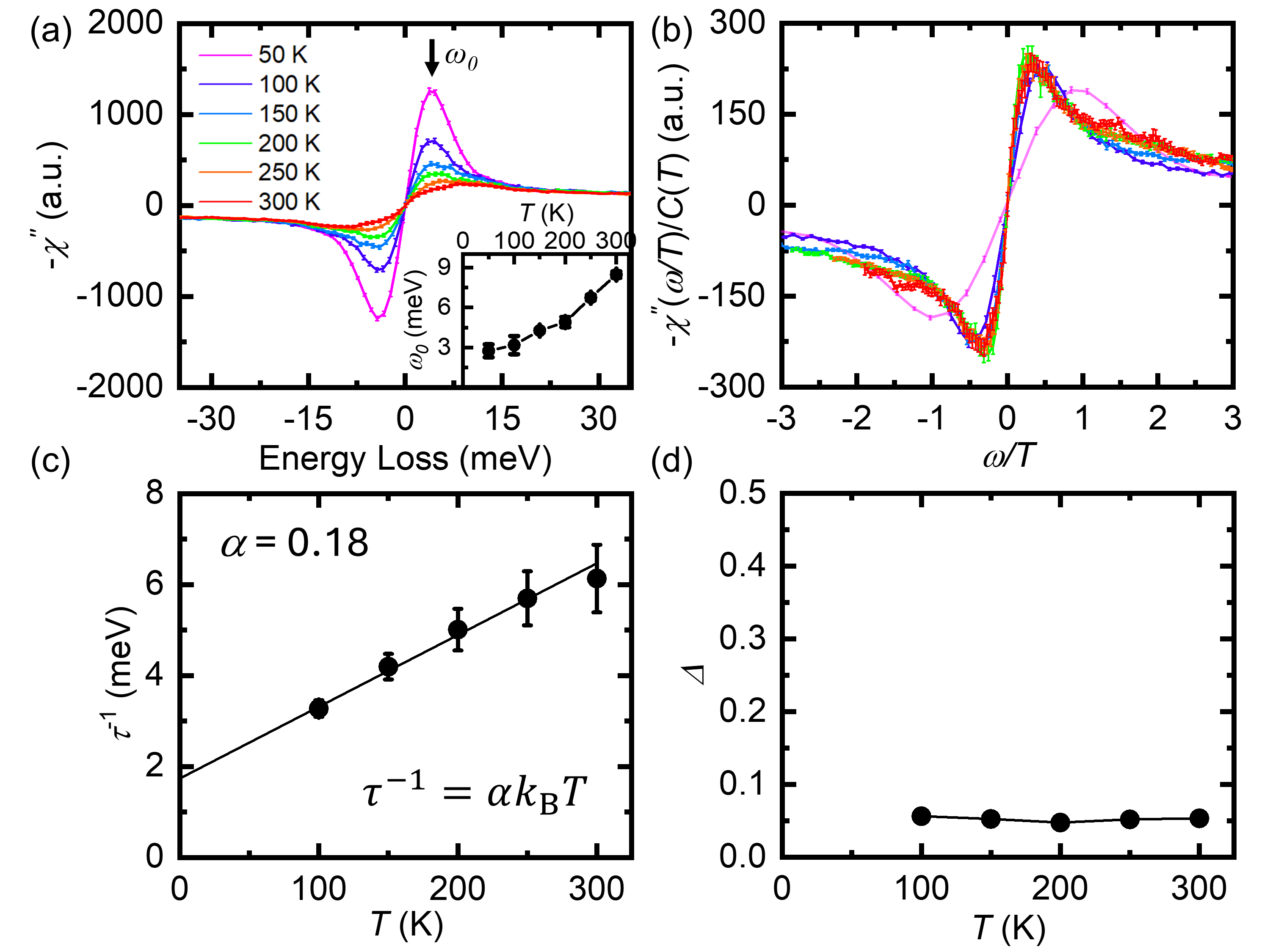}
    \caption{\textbf{Temperature-dependent quasielastic response at $q=0.36$ r.l.u. on Sample 2, showing our data are reproducible.} (a) Density response shows a peak at $\omega_0$ similar to Fig. \ref{q_36}(b). The inset illustrates the linear-in-$T$ behavior which curves at low temperatures. (b) $\omega/T$ ``data collapse” at different temperatures comparable to Fig. \ref{q_36}(c), with the $T=50$ K data influenced by resolution and superconductivity effects. (c) Dissipation in the relaxational dynamics shows a similar behavior to Fig. \ref{fits}(k), with similar value of $\alpha=0.18$. (d) The conformal dimension is $\Delta=0.05$, consistent with Fig. \ref{fits}(l).}
    \label{reproduce_scale}
\end{figure}

\begin{figure}[H]
    \centering
    \includegraphics[scale=0.47]{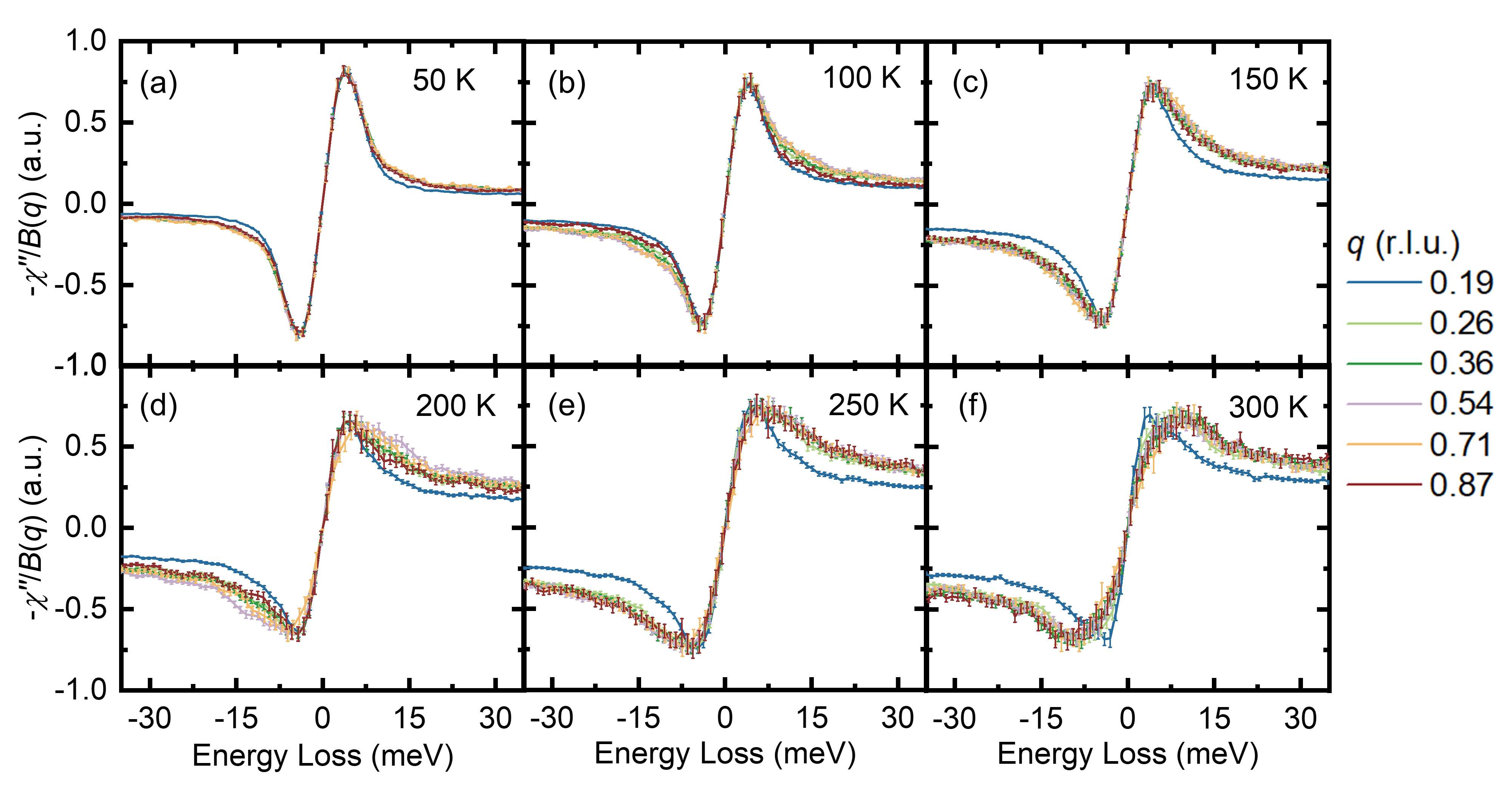}
    \caption{\textbf{Momentum-dependence of susceptibility at different temperatures on Sample 2.} (a)-(f) The susceptibility is divided by $B(q)$ to match the peak height at $\omega_0$ for 50 K $< T < $300 K. The susceptibility is \textit{separable}, similar to Fig. \ref{momentum}.}
    \label{reproduce_momentum}
\end{figure}

\bibliography{bibliography}

\begin{thebibliography}{10}
\providecommand{\url}[1]{{#1}}
\providecommand{\urlprefix}{URL }
\providecommand{\doi}[1]{\url{https://doi.org/#1}}
\bibcommenthead

\bibitem{Hussey2004}
N.E. Hussey, K.~Takenaka, H.~Takagi, Universality of the {Mott–Ioffe–Regel} limit in metals.
\newblock Philos. Mag. \textbf{84}, 2847--2864 (2004)

\bibitem{Ashcroft76}
N.W. Ashcroft, N.D. Mermin, \emph{{S}olid {S}tate {P}hysics} (Holt-Saunders, 1976)

\bibitem{basov2005electrodynamics}
D.N. Basov, T.~Timusk, Electrodynamics of {high-$T_c$} superconductors.
\newblock Rev. Mod. Phys. \textbf{77}, 721--779 (2005)

\bibitem{marel2003quantum}
D.~van~der Marel, H.J.A. Molegraaf, J.~Zaanen, Z.~Nussinov, F.~Carbone, A.~Damascelli, H.~Eisaki, M.~Greven, P.H. Kes, M.~Li, Quantum critical behaviour in a {high-$T_c$} superconductor.
\newblock Nature \textbf{425}, 271--274 (2003)

\bibitem{reber2019unified}
T.J. Reber, et~al., A unified form of low-energy nodal electronic interactions in hole-doped cuprate superconductors.
\newblock Nat. Commun. \textbf{10}, 5737 (2019)

\bibitem{Vishik2010}
I.M. Vishik, W.S. Lee, R.H. He, M.~Hashimoto, Z.~Hussain, T.P. Devereaux, Z.X. Shen, {ARPES} studies of cuprate {Fermiology}: superconductivity, pseudogap and quasiparticle dynamics.
\newblock New J. Phys. \textbf{12}, 105008 (2010)

\bibitem{zaanen2019planckian}
J.~Zaanen, Planckian dissipation, minimal viscosity and the transport in cuprate strange metals.
\newblock SciPost Phys. \textbf{6}, 061 (2019)

\bibitem{chaikin}
P.M. Chaikin, T.C. Lubensky, \emph{Principles of Condensed Matter Physics} (Cambridge University Press, Cambridge, 1995)

\bibitem{goldenfeld}
N.~Goldenfeld, \emph{Lectures On Phase Transitions And The Renormalization Group (1st ed.)} (CRC Press, Boca Raton, 1992)

\bibitem{varma1989phenomenology}
C.M. Varma, P.B. Littlewood, S.~Schmitt-Rink, E.~Abrahams, A.E. Ruckenstein, Phenomenology of the normal state of {Cu-O} high-temperature superconductors.
\newblock Phys. Rev. Lett. \textbf{63}, 1996--1999 (1989)

\bibitem{PinesNozieres1973}
D.~Pines, P.~Nozi\`eres, \emph{The Theory of Quantum Liquids} (Perseus Books, Cambridge, MA, 1999)

\bibitem{mitrano2018anomalous}
M.~Mitrano, et~al., Anomalous density fluctuations in a strange metal.
\newblock Proc. Natl. Acad. Sci. USA \textbf{115}, 5392--5396 (2018)

\bibitem{SachdevYe1993}
S.~Sachdev, J.~Ye, Gapless spin-fluid ground state in a random quantum {Heisenberg} magnet.
\newblock Phys. Rev. Lett. \textbf{70}, 3339--3342 (1993)

\bibitem{subir1}
A.A. Patel, H.~Guo, I.~Esterlis, S.~Sachdev, Universal theory of strange metals from spatially random interactions.
\newblock Science \textbf{381}, 790--793 (2023)

\bibitem{subir2}
C.~Li, D.~Valentinis, A.A. Patel, H.~Guo, J.~Schmalian, S.~Sachdev, I.~Esterlis, Strange metal and superconductor in the two-dimensional {Yukawa-Sachdev-Ye-Kitaev} model.
\newblock Phys. Rev. Lett. \textbf{133}, 186502 (2024)

\bibitem{subir3}
A.A. Patel, P.~Lunts, S.~Sachdev, Localization of overdamped bosonic modes and transport in strange metals.
\newblock Proc. Natl. Acad. Sci. USA \textbf{121}, e2402052121 (2024)

\bibitem{polyakov1}
A.M. Polyakov, Conformal symmetry of critical fluctuations.
\newblock JETP Lett. \textbf{12}, 381--383 (1970)

\bibitem{polyakov2}
A.A. Belavin, A.M. Polyakov, A.B. Zamolodchikov, Infinite conformal symmetry in two-dimensional quantum field theory.
\newblock Nucl. Phys. B \textbf{241}, 333--380 (1984)

\bibitem{Tsvelik1997}
M.C. Aronson, M.B. Maple, P.~de~Sa, A.M. Tsvelik, R.~Osborn, Non–fermi-liquid scaling in {UCu$_{5-x}$Pd$_x$} ($x=1,1.5$): A phenomenological description.
\newblock Europhys. Lett. \textbf{40}, 245--250 (1997)

\bibitem{schulte2002interplay}
K.H.G. Schulte, The interplay of spectroscopy and correlated materials.
\newblock Ph.D. thesis, University of Groningen (2002)

\bibitem{Chen2024}
J.~Chen, et~al., Consistency between reflection momentum-resolved electron energy loss spectroscopy and optical spectroscopy measurements of the long-wavelength density response of {${\mathrm{Bi}}_{2}{\mathrm{Sr}}_{2}{\mathrm{CaCu}}_{2}{\mathrm{O}}_{8+x}$}.
\newblock Phys. Rev. B \textbf{109}, 045108 (2024)

\bibitem{husain2019crossover}
A.A. Husain, et~al., Crossover of charge fluctuations across the strange metal phase diagram.
\newblock Phys. Rev. X \textbf{9}, 041062 (2019)

\bibitem{ARCMP}
P.~Abbamonte, J.~Fink, Collective charge excitations studied by electron energy-loss spectroscopy {\href{https://arxiv.org/abs/2404.04670}{{arXiv:2404.04670}}}

\bibitem{vig2017measurement}
S.~Vig, et~al., Measurement of the dynamic charge response of materials using low-energy, momentum-resolved electron energy-loss spectroscopy {(M-EELS)}.
\newblock SciPost Phys. \textbf{3}, 026 (2017)

\bibitem{wen2008}
J.S. Wen, Z.J. Xu, G.Y. Xu, M.~Hücker, J.M. Tranquada, G.D. Gu, Large {Bi-2212} single crystal growth by the floating-zone technique.
\newblock J. Cryst. Growth \textbf{310}, 1401--1404 (2008)

\bibitem{phelps1993}
R.B. Phelps, P.~Akavoor, L.L. Kesmodel, J.E. Demuth, D.B. Mitzi, Surface phonons on {Bi$_2$Sr$_2$CaCu$_2$0$_{8+\delta}$}.
\newblock Phys. Rev. B \textbf{48}, 12936--12940 (1993)

\bibitem{persson1990}
B.N.J. Persson, J.E. Demuth, High-resolution electron-energy-loss study of the surfaces and energy gaps of cleaved high-temperature superconductors.
\newblock Phys. Rev. B \textbf{42}, 8057--8072 (1990)

\bibitem{demuth1990}
J.E. Demuth, B.N.J. Persson, F.~Holtzberg, C.V. Chandrasekhar, Surface and superconducting properties of cleaved high-temperature superconductors.
\newblock Phys. Rev. Lett. \textbf{64}, 603--606 (1990)

\bibitem{Egerton2011}
R.F. Egerton, \emph{Electron Energy-Loss Spectroscopy in the Electron Microscope} (Springer, New York Dordrecht Heidelberg London, 2011)

\bibitem{Husain2023}
A.A. Husain, et~al., Pines’ demon observed as a {3D} acoustic plasmon in {Sr$_2$RuO$_4$}.
\newblock Nature \textbf{621}, 66--70 (2023)

\bibitem{Li2023}
J.~Li, J.~Li, J.~Tang, Z.~Tao, S.~Xue, J.~Liu, H.~Peng, X.Q. Chen, J.~Guo, X.~Zhu, Direct observation of topological phonons in graphene.
\newblock Phys. Rev. Lett. \textbf{131}, 116602 (2023)

\bibitem{Kogar2017}
A.~Kogar, et~al., Signatures of exciton condensation in a transition metal dichalcogenide.
\newblock Science \textbf{358}, 1314--1317 (2017)

\bibitem{Diaconescu2007}
B.~Diaconescu, et~al., Low-energy acoustic plasmons at metal surfaces.
\newblock Nature \textbf{448}, 57--59 (2007)

\bibitem{faulkner2}
T.~Faulkner, G.T. Horowitz, M.M. Roberts, Holographic quantum criticality from multi-trace deformations.
\newblock JHEP \textbf{04}, 051 (2011)

\bibitem{faulkner1}
T.~Faulkner, H.~Liu, J.~McGreevy, D.~Vegh, Emergent quantum criticality, {Fermi} surfaces, and {${\mathrm{AdS}}_{2}$}.
\newblock Phys. Rev. D \textbf{83}, 125002 (2011)

\bibitem{Gori3dCI}
G.~Gori, A.~Trombettoni, Conformal invariance in three dimensional percolation.
\newblock J. Stat. Mech. \textbf{1507}, P07014 (2015)

\bibitem{Rattazzi}
R.~Rattazzi, V.S. Rychkov, E.~Tonni, A.~Vichi, Bounding scalar operator dimensions in {4$D$ CFT}.
\newblock JHEP \textbf{12}, 031 (2008)

\bibitem{polchinski}
J.~Polchinski, Scale and conformal invariance in quantum field theory.
\newblock Nucl. Phys. B \textbf{303}, 226--236 (1988)

\end{thebibliography}

\end{document}